\documentclass{article}

\usepackage{spconf,amsmath,graphicx}
\usepackage{mathtools}
\usepackage{amssymb}
\usepackage{amsmath}
\usepackage[utf8]{inputenc}
\usepackage{algpseudocode}
\usepackage{array}

\usepackage{tikz}
\usepackage{pgfplots}
\usetikzlibrary{shapes,snakes}

\newcommand{\argmin}{\operatornamewithlimits{argmin}}

\newcommand{\softth}{\operatornamewithlimits{SoftThreshold}}

\graphicspath{{pictures/}}

\def\XX{{Multi-SSSA}}
\def\muu{multi-dimensional}

\title{Multi-dimensional sparse structured signal approximation\\ using split Bregman iterations}
\name{Yoann Isaac$^{1,2}$, Quentin Barthélemy$^{1}$, Jamal Atif$^{2}$, Cédric Gouy-Pailler$^{1}$, Michèle Sebag$^{2}$ \thanks{The work presented in this paper has been partially funded by DIGITEO under the grant 2011-053D.}}
\address{
         \begin{tabular}{c c c}
         $^1$ CEA, LIST & \hspace{2cm} & $^2$ TAO, CNRS $-$ INRIA $-$ LRI\\
         Data Analysis Tools Laboratory & & Université Paris-Sud\\
         91191 Gif-sur-Yvette CEDEX, FRANCE & & 91405 Orsay, FRANCE \\
         \end{tabular}
        }

\begin{document}
%
\maketitle
\arraycolsep=3pt
\begin{abstract}

The paper focuses on the sparse approximation of signals using overcomplete 
representations, such that it preserves the (prior) structure of
multi-dimensional signals. The underlying optimization problem is tackled
using a multi-dimensional split Bregman optimization
approach. An extensive 
empirical evaluation shows how the proposed approach compares to the state of
the art depending on
the signal features. \\
\end{abstract}
%

\section{Introduction}
\label{sec:intro}

Dictionary-based representations proceed by approximating a signal via a
linear combination of dictionary elements, referred to as atoms. Sparse
dictionary-based representations, where each signal involves few atoms,
have been thoroughly investigated for their good
properties, as they enable robust transmission (compressed sensing
\cite{donoho2006compressed}) or image in-painting \cite{mairal2008sparse}. The
dictionary is either given, based on the domain knowledge, or learned from the
signals \cite{tosic2011dictionary}.

The so-called sparse approximation algorithm aims at finding a
sparse approximate representation of the considered signals using this
dictionary, by minimizing a weighted sum of the approximation loss and the
representation sparsity (see \cite{rakotomamonjy2011surveying} for a survey).
When available, prior knowledge about the application
domain can also be used to guide the search toward ``plausible'' decompositions.\\
This paper focuses on sparse approximation enforcing a structured decomposition property, defined as follows. Let the signals be structured (e.g. being recorded in consecutive time steps); the structured decomposition property then requires that the signal structure is preserved in the dictionary-based representation (e.g. the atoms involved in the approximation of consecutive signals
have ``close'' weights). The structured decomposition property is enforced through adding a total variation (TV) penalty to the minimization objective.

In the 1D case, the minimization of the above overall objective can be tackled using the fused-LASSO approach first introduced in \cite{tibshirani2005sparsity}. In the case of multi-dimensional (also called multi-channel)
signals\footnote{Our motivating application considers electro-encephalogram (EEG) signals, where the number of sensors ranges up to a few tens.}
however, the minimization problem presents additional difficulties.
The first contribution of the paper is to show how this problem can be handled
efficiently, by extending the (mono-dimensional) split Bregman fused-LASSO solver presented
in \cite{ye2011split}, to the multi-dimensional case.
The second contribution is a comprehensive experimental study, comparing
state-of-the-art algorithms
to the presented approach referred to as \XX\ and establishing their relative
performance depending on
diverse features of the structured signals. 

This paper is organized as follows. The Section~\ref{sec:prob} introduces the formal background. 
The proposed optimization approach is described in Section~\ref{sec:optim}. 
Section \ref{sec:experiments} presents our experimental settings
and reports on the results. The presented approach is discussed w.r.t. related
work in Section~\ref{sec:rel} and the paper concludes with some perspectives
for further researches.


\section{Problem statement}
\label{sec:prob}

Let $Y=[{\bf y}_1, \dots, {\bf y}_T] \in \mathbb{R}^{C \times T}$ be a matrix made of $T$ $C$-dimensional signals, and $\Phi \in \mathbb{R}^{C \times N}$ an overcomplete dictionary of $N$ normalized atoms ($N>C$).
We consider the linear model:
\begin{eqnarray}
{\bf y}_t = \Phi {\bf x}_t + {\bf e}_t,~~t \in \{1, \dots, T\} \enspace \nonumber,
\end{eqnarray}
in which $X=[{\bf x}_1, \dots, {\bf x}_T] \in \mathbb{R}^{N \times T}$ stands
for the decomposition matrix and $E=[{\bf e}_1, \dots, {\bf e}_T] \in \mathbb{R}^{C \times T}$ is a Gaussian noise matrix. \noindent The sparse structured decomposition problem consists of
approximating the ${\bf y}_t$, $t \in \{1, \dots, T\}$, by decomposing them on
the dictionary $\Phi$, such that the structure of the 
decompositions ${\bf x}_t$ reflects that of the signals ${\bf y}_t$. This goal
is formalized as the minimization of the objective function:\footnote{$\|A\|_p = (\sum_{i}\sum_{j}
|A_{i,j}|^p)^{\frac{1}{p}}$. The case $p=2$ corresponds to the classical
Frobenius norm.} 
\begin{eqnarray}
\min_{X} \|Y-\Phi X\|_2^2 + \lambda_1 \|{X}\|_1 + \lambda_2 \|{X} P\|_1 
\,\enspace ,
\end{eqnarray}
\noindent where  $\lambda_1$ and $\lambda_2$ are regularization coefficients
and $P$ encodes the
signal structure (provided by the prior knowledge) as in \cite{chen2010graph}.
In the remainder of the paper, the considered structure is that of the
temporal ordering of the signals, {\em i.e.}
$\|{X} P\|_1 = \sum_{t=2}^{T}\|X_{t}-X_{t-1}\|_1$.


\section{Optimization strategy}
\label{sec:optim}

\subsection{Algorithm description}

Bregman iterations have shown to be very efficient for $\ell_1$ regularized problems~\cite{goldstein2009split}. For convex problems with linear constraints, the split Bregman iteration
technique is equivalent to the method of multipliers and the augmented
Lagrangian one \cite{wu2010augmented}.
The iteration scheme presented in \cite{ye2011split} considers an augmented
Lagrangian formalism. We have chosen here to present ours with the initial
split Bregman formulation.\\

First, let us restate the sparse approximation problem: 
\begin{eqnarray}
\begin{array}{ll}
\label{eqn:base}
\min_{X,A,B} &\|Y-\Phi X\|_2^2 + \lambda_1 \|A\|_1 + \lambda_2 \|B\|_1\\
\text{s.t.} ~~ &A = X  \\
       &B = X P 
\end{array} \enspace .
\end{eqnarray}
This reformulation is a key step of the split Bregman method. It decouples the three terms and allows to optimize them separately within the Bregman iterations. To set-up this iteration scheme, Eq.~(\ref{eqn:base}) must be transform to an unconstrained problem:
\begin{eqnarray}
\begin{array}{ll}
\min_{X, A, B} &\|Y-\Phi X\|_2^2 + \lambda_1 \|A\|_1 + \lambda_2 \|B\|_1 \\
&+ \frac{\mu_1}{2} \|X - A\|_2^2 + \frac{\mu_2}{2} \|X P - B\|_2^2 \nonumber
\end{array} \enspace .
\end{eqnarray}

\noindent The split Bregman scheme could then be expressed as~\cite{goldstein2009split}:
\begin{eqnarray}
\label{eqn:primal}
(X^{i+1}, A^{i+1}, B^{i+1}) = & \argmin_{X, A, B} \|Y-\Phi X\|_2^2\nonumber\\
&+ \lambda_1 \|A\|_1 + \lambda_2 \|B\|_1 \\
& + \frac{\mu_1}{2} \|X - A + D_{A}^i\|_2^2 \nonumber\\
&+ \frac{\mu_2}{2} \|X P - B + D_{B}^i\|_2^2 \nonumber\\
D_{A}^{i+1} = & D_{A}^{i} + (X^{i+1} - A^{i+1})\nonumber\\
D_{B}^{i+1} = & D_{B}^{i} + (X^{i+1} P - B^{i+1}) \nonumber \enspace .
\end{eqnarray}

\noindent Thanks to the split of the three terms, the
minimization of Eq.~(\ref{eqn:primal}) could be performed iteratively by
alternatively updating variables in the system:
\begin{eqnarray}
	\label{eqn:theta_update}
	X^{i+1}\!\!\! &=\! \argmin_{X} \|Y-\Phi X\|_2^2 + \frac{\mu_1}{2} \|X - A^i + D_{A}^i\|_2^2 \nonumber \\
	& + \frac{\mu_2}{2} \|X P - B^i + D_{B}^i\|_2^2
	\\
	\label{eqn:A_update}
	A^{i+1}\!\!\! &=\! \argmin_{A} \lambda_1 \|A\|_1 + \frac{\mu_1}{2} \|X^{i+1} - A + D_{A}^i\|_2^2
	\\
	\label{eqn:B_update}
	B^{i+1}\!\!\! &=\! \argmin_{B} \lambda_2 \|B\|_1 + \frac{\mu_2}{2} \|X^{i+1} P - B + D_{B}^i\|_2^2 
\end{eqnarray}

\noindent Only few iterations of this system are necessary for convergence. In our implementation,  this update is only performed once at each iteration of the global optimization algorithm.\\

\noindent Eq.~(\ref{eqn:A_update}) and Eq.~(\ref{eqn:B_update}) could be resolved with the soft-thresholding operator:
\begin{eqnarray}
	A^{i+1} = & \hspace{-0.5cm} \softth_{\frac{\lambda_1}{\mu_1}\|.\|_1} (X^{i+1}  +  D_{A}^i)
	\\
	B^{i+1} = & \softth_{\frac{\lambda_2}{\mu_2} \|.\|_1} (X^{i+1} P  + D_{B}^i) \enspace .
\end{eqnarray}

\noindent Solving Eq.~(\ref{eqn:theta_update}) requires the minimization of a
convex differentiable function which can be performed via classical
optimization methods. We propose here to solve it deterministically. The main
difficulty in extending \cite{ye2011split} to the \muu\ signals case rely on
this step. Let us define $H$ from Eq.~(\ref{eqn:theta_update}) such as:
\begin{eqnarray}
X^{i+1} = &\argmin_{X} H(X) \nonumber \enspace .
\end{eqnarray}
Differentiating this expression with respect to $X$ yields:
\begin{eqnarray}
\frac{d}{dX}H = &(2\Phi^T \Phi + \mu_1 I)X + X (\mu_2 PP^T) - 2\Phi Y \\&+
\mu_1(D_{A}^i - A^i) + \mu_2(D_{B}^i - B^i)P^T \nonumber \enspace ,
\end{eqnarray} 
\noindent where $I$ is the identity matrix. The minimum $\hat{X}=X^{i+1}$ of Eq.~(\ref{eqn:theta_update}) is obtained by solving $\frac{d}{dX}H(\hat{X}) = 0$ which is a Sylvester equation:
\begin{eqnarray}
\label{eqn:syl}
W\hat{X} + \hat{X} Z = M^i \enspace ,
\end{eqnarray}
with $W=2\Phi^T \Phi + \mu_1 I$, $Z=\mu_2 PP^T$ and $M=- D_{A}^i + 2\Phi^T Y + \mu_1 A^i + (\mu_2 B^i - D_{B}^i)P^T$. Fortunately, in our case, $W$ and $Z$ are real symmetric matrices. Thus, they can be diagonalized as follows:
\begin{eqnarray}
W = F D_w F^T \nonumber\\
Z = G D_z G^T \nonumber
\end{eqnarray}
where $F$ and $G$ are orthogonal matrices. Eq.~(\ref{eqn:syl}) becomes:
\begin{eqnarray}
& D_w\hat{X}{} ' + \hat{X}{}' D_z = M^{i}{}' \enspace ,
\end{eqnarray}
with $\hat{X}{}'= F^T \hat{X} G$ and $M^{i}{}' = F^T M^{i} G$. \\
$\hat{X}{}'$ is then obtained by:
\begin{eqnarray}
\forall t \in \{1, \dots, T\}~~\hat{X}{}'(:, t) = (D_w + D_z(t,t)I)^{-1} M^{i}{}'(:, t) \nonumber
\end{eqnarray}
where the notation $(:,t)$ indices the column $t$ of matrices. Going back to $\hat{X}$ could be performed with: $\hat{X} = F\hat{X}{}' G^T$.\\
$W$ and $Z$ being independent of the iteration ($i$) considered, their diagonalizations are done only once and for all as well as the computation of the terms $(D_w + D_z(t,t)I)^{-1}$, $\forall t \in \{1, \dots, T\}$. Thus, this update does not require heavy computations. The full algorithm is summarized below.

\subsection{\XX~sum up}
{
\ninept
Inputs: $Y$, $\Phi$, $P$. ~~ Parameters: $\lambda_1$, $\lambda_2$, $\mu_1$, $\mu_2$, $\epsilon$, $iterMax$, $kMax$
\begin{algorithmic}[1]
\State Init $D_A^0$, $D_B^0$, $X^0$ and set $B^0 = X^0 P$,~ $A^0=X^0$, 
\State $W=2\Phi^T \Phi + \mu_1 I$ and $Z=\mu_2PP^T$.
\State Compute $D_w$, $D_z$, $F$ and $G$ from $W$ and $Z$.
\State Precompute ($t \to T$), $D_{temp}^t = (D_y + D_z(t,t)I)^{-1}$.
\State $i = 0$
\While{$i \leq iterMax $ and $\frac{\|X^{i}- X^{i-1}\|_2}{\|X^{i}\|_2} \geq \epsilon $}
	\State $k=0$
	\State $X^{temp} = X^{i}$; $A^{temp} = A^{i}$; $B^{temp} = B^{i}$
	\For{$k \to kMax$}
		\State $M' = F^T (2\Phi^T Y - \mu_1 (D_A^i-A^{temp}) - \mu_2 (D_B^i-B^{temp})P^T) G$
		\For{$t \to T$}
			\State $X^{temp}(:, t) = D_{temp}^t M'(:, t)$
		\EndFor
		\State $X^{temp} = FX^{temp} G^T$
		\State $A^{temp} = \softth_{\frac{\lambda_1}{\mu_1}\|.\|_1}(X^{temp}  + D_A^i)$
		\State $B^{temp} = \softth_{\frac{\lambda_2}{\mu_2}\|.\|_1}(X^{temp} P  + D_B^i)$
	\EndFor
	\State $X^{i+1} = X^{temp}$; $A^{i+1} = A^{temp}$; $B^{i+1} = B^{temp}$ 
	\State $D_A^{i+1} = D_A^i + (X^{i+1} - A^{i+1})$
 	\State $D_B^{i+1} = D_B^i + (X^{i+1} P - B^{i+1})$
    \State $i=i+1$
\EndWhile
\end{algorithmic}
}


\section{Experimental evaluation}
\label{sec:experiments}
The following experiment aims at assessing the efficiency of our approach in decomposing signals built with particular regularities. We compare it both to algorithms coding each signal separately, the orthogonal matching pursuit (OMP) \cite{pati1993orthogonal} and the LARS \cite{efron2004least} (a LASSO solver), and to methods performing the decomposition simultaneously, the simultaneous OMP (SOMP) and an proximal method solving the group-LASSO problem (FISTA~\cite{beck2009fast}).

\subsection{Data generation}

From a fixed random overcomplete dictionary $\Phi$, a set of $K$ signals having piecewise constant structures have been created. Each signal $Y$ is synthesized from the dictionary $\Phi$ and a built decomposition matrix $X$ with $Y=\Phi X$\\
The TV penalization of the fused-LASSO regularization makes him more suitable to deal with data having abrupt changes. Thus, the decomposition matrices of signals have been built as linear combinations of specific activities which have been generated as follows:
\begin{eqnarray}
P_{ind, m, d}(i,j) = \left\{
\begin{array}{ll} \nonumber
	0 & \text{if} \ i \neq ind\\
	{\cal H}(j-(m-\frac{d \times T}{2})) \\
	- {\cal H}(j-(m+\frac{d \times T}{2}))\ \ \ \ & \text{if} \ i = k
\end{array}
\right.
\end{eqnarray}
where $P \in \mathbb{R}^{N \times T}$, ${\cal H}$ is the Heaviside function, $ind \in \{1, \dots, N\}$ is the index of an atom, $m$ is the center of the activity and $d$ its duration. Each decomposition matrix $X$ could then be written:
\begin{eqnarray}
X = \sum_{i=1}^{n_a} a_i P_{ind_i, m_i, d_i} \ , \nonumber
\end{eqnarray}
where $n_a$ is the number of activities appearing in one signal and the $a_i$ stand for the activation weights. 
An example of such signal is given in the Figure \ref{fig:data_synth} below.\\   
 
\begin{figure}[h!]
\begin{center}
\begin{tabular}{lll}
\includegraphics[scale=0.17]{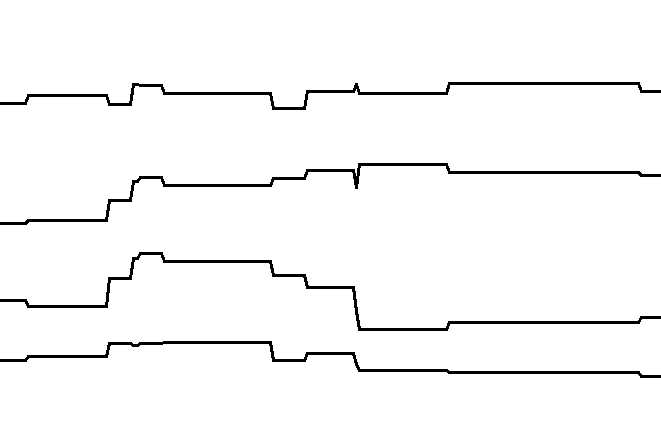}&
\includegraphics[scale=0.4]{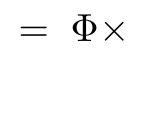}
\includegraphics[scale=0.17]{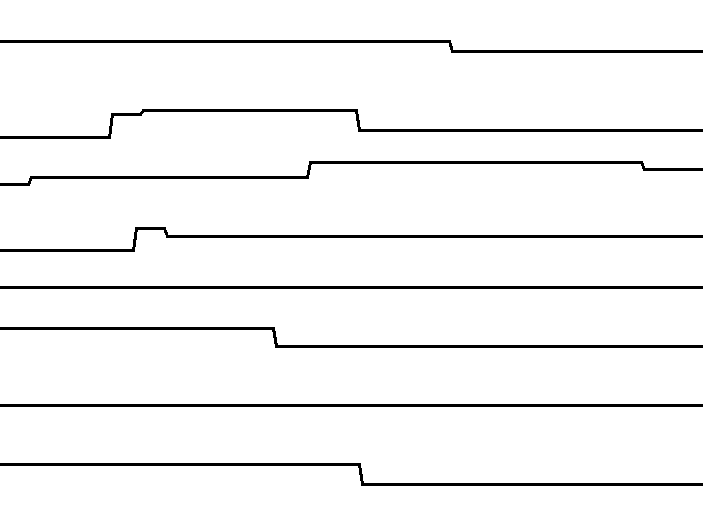}
\end{tabular}
\end{center}
\caption{Built signal, with $C=4$ channels and $N=8$ atoms.}
\label{fig:data_synth}
\end{figure}

\subsection{Experimental setting}

Each method has been applied to the previously created signals. Then the distances between the estimated decomposition matrices $\hat{X}$ and the real ones $X$ have been calculated as follows:     
\begin{eqnarray}
dist(X, \hat{X}) = \frac{\|X - \hat{X}\|_{2}}{\| X \|_{2}} \ \nonumber.
\end{eqnarray}

The goal was to understand the influence of the number of activities ($n_a$) and the range of durations ($d$) on the efficiency of the fused-LASSO regularization compared to others sparse coding algorithms. The scheme of experiment described above has been carried out with the following grid of parameters:
\begin{itemize}
 \item $n_a \in \{20, 30, \dots, 110\}$,
 \item $d\sim \mathcal{U}(d_{min}, d_{max})$ \\
 $(d_{min} d_{max}) \in \{(0.1, 0.15), (0.2, 0.25), \dots, (1, 1)\}$ 
\end{itemize}
For each point in the above parameter grid, two sets of signals has been created: a train set allowing to determine for each method the best regularization coefficients and a test set designed for evaluate them with these coefficients.

\noindent Other parameters have been chosen as follows:
\begin{center}
    \begin{tabular}{ll}
    \hline
    Model & Activities\\
    \hline
    $C=20$&$m\sim \mathcal{U}(0,T)$\\
    $T=300$&$a\sim \mathcal{N}(0, 2)$\\
    $N=40$&$ind\sim \mathcal{U}(1, N)$\\
    $K=100$&\\
    \hline
    \end{tabular}
\end{center}

\noindent Dictionaries have been randomly generated using Gaussian independent distributions on individual elements and present low coherence.

\subsection{Results and discussion}
\begin{figure*}
\centering
\includegraphics[scale=0.15]{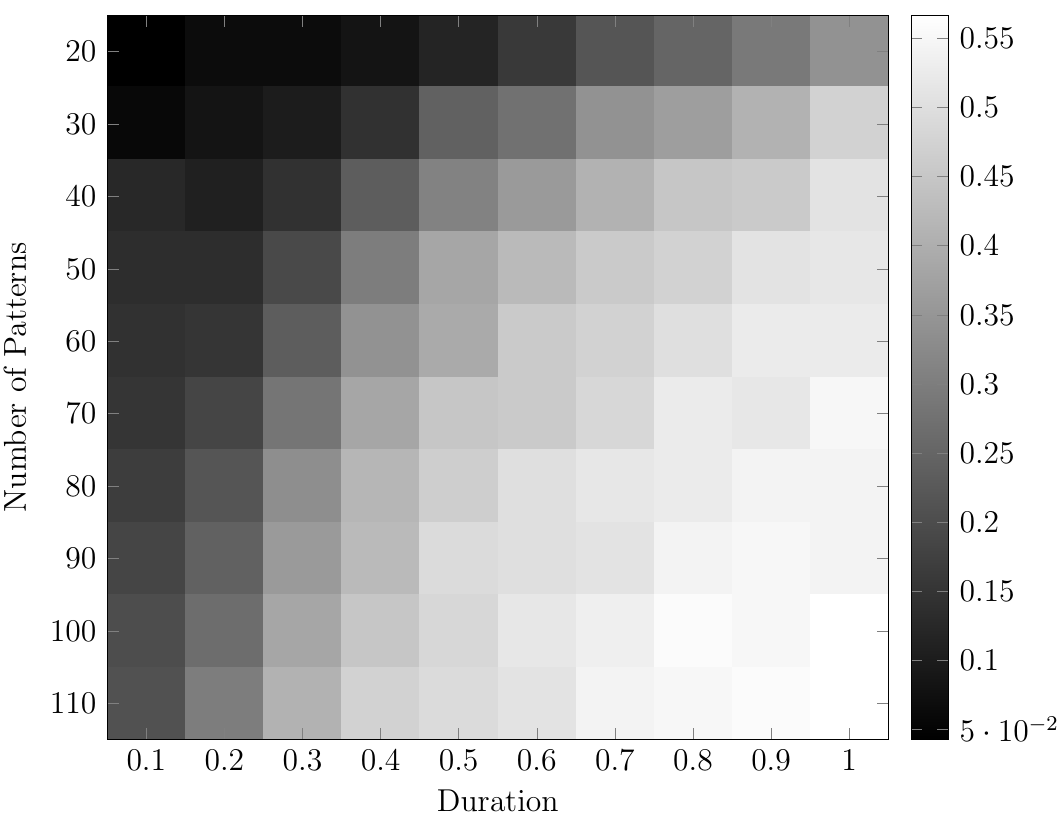}
\hspace{0.3cm}
\includegraphics[scale=0.15]{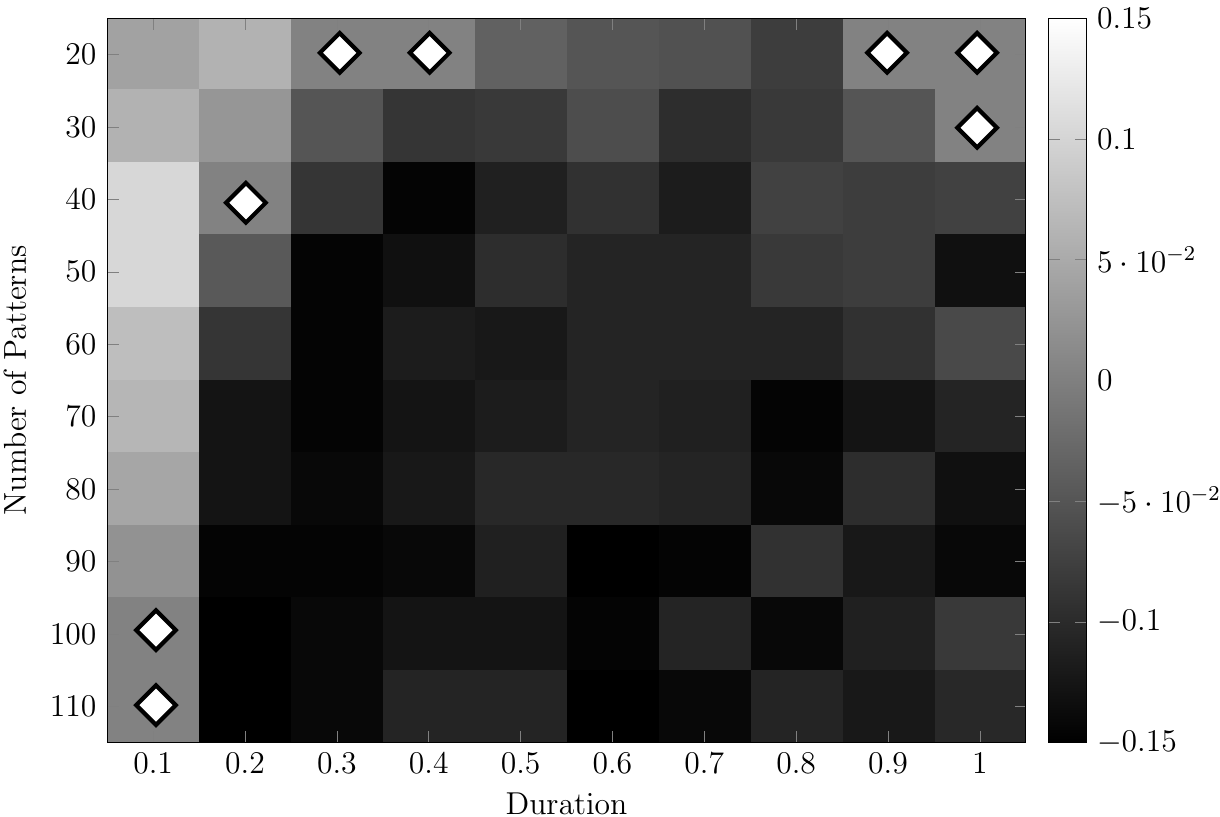}
\hspace{0.3cm}
\includegraphics[scale=0.15]{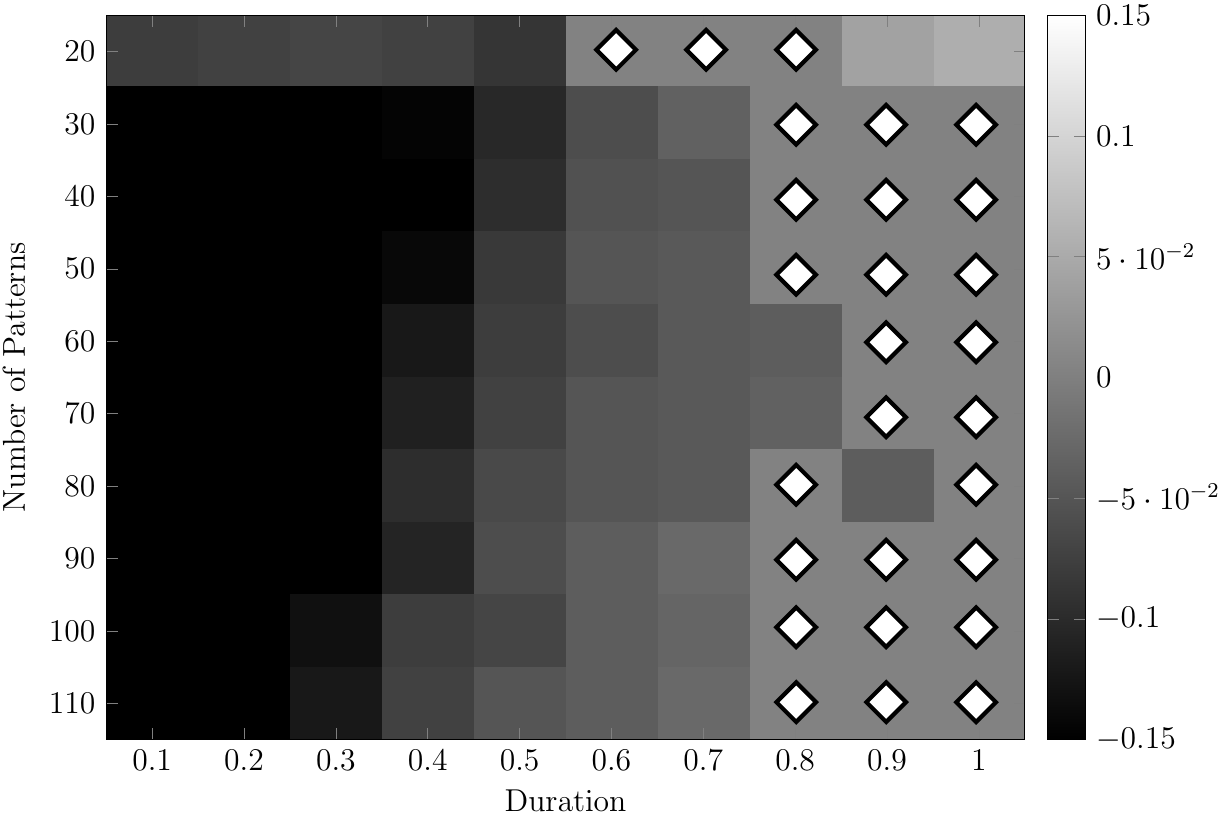}

\caption{Left: Mean distances $dist$ obtained with the \XX. Middle: Difference between the mean distances obtained with the \XX~and those obtained with the LARS. Right: Difference between the mean distances obtained with the \XX~and those obtained with the Group-LASSO solver. The white diamonds correspond to non-significant differences between the means distances.}
\label{fig:valid}
\end{figure*}

In order to evaluate the proposed algorithm, for each point $(i,j)$ in the above grid of parameters, the mean (among test signals) of the previously defined distance $dist$ has been computed for each method and compared to the mean obtained by the \XX. A paired t-test ($p<0.05$) has then been performed to check the significance of these differences.\\
Results are displayed in Figure~\ref{fig:valid}. In the ordinate axis, the number of patterns increases from the top to the bottom and in the abscissa axis, the duration grows from left to right. The left image displays the mean distances obtained by the \XX. Unsurprisingly, the difficulty of finding the ideal decomposition increases with the number of patterns and their durations. The middle and right figures present its performances compared to other methods by displaying the differences (point to point) of mean distances in grayscale. These differences are calculated such that, negative values (darker blocks) means that our method outperform the other one. The white diamonds correspond to non-significant differences of mean distances.  Results of the OMP and the LARS are very similar as well as those of the SOMP and the group-LASSO solver. We only display here the matrices comparing our method to the LARS and the group-LASSO solver.

Compared to the OMP and the LARS, our method obtains same results as them when only few atoms are active at the same time. It happens in our artificial signals when only few patterns have been used to create decomposition matrices and/or when the pattern durations are small. On the contrary, when many atoms are active simultaneously, the OMP and LARS are outperformed by the above algorithm which use inter-signal prior information to find better decompositions.\\
Compared to the SOMP and the group-LASSO solver, results depend more on the duration of patterns. When patterns are long and not too numerous, theirs performances is similar to the fused-LASSO one. The SOMP is outperformed in all other cases. On the contrary, the group-LASSO solver is outperformed only when patterns have short/medium durations.\\


\section{Relation to prior works}
\label{sec:rel}
The simultaneous sparse approximation of \muu\ signals has been widely studied during these last years~\cite{chen2006theoretical} and numerous methods developed~\cite{tropp2006algorithms1, tropp2006algorithms2, gribonval2008atoms, cotter2005sparse, rakotomamonjy2011surveying}. More recently, the concept of structured sparsity has considered the encoding of priors in complex regularizations~\cite{huang2011learning, jenatton00377732}. Our problem belongs to this last category with a regularization combining a classical sparsity term and a Total Variation one. This second term has been studied intensively for image denoising as in the ROF model~\cite{rudin1992nonlinear, darbon2005fast}.\\
The combination of these terms has been introduced as the fused-LASSO \cite{tibshirani2005sparsity}. Despite its convexity, the two $\ell_1$ non-differentiable terms make it difficult to solve. The initial paper~\cite{tibshirani2005sparsity} transforms it to a quadratic problem and uses standard optimization tools (SQOPT). Increasing the number of variables, this approach can not deal with large-scale problems. A path algorithm has been developed but is limited to the particular case of the fused-LASSO signal approximator~\cite{hoefling2010path}. More recently, scalable approaches based on proximal sub-gradient methods \cite{liu2010efficient}, ADMM \cite{wahlberg2012admm} and split Bregman iterations \cite{ye2011split} have been proposed for the general fused-LASSO.\\
To the best of our knowledge, the \muu~fused-LASSO in the context of overcomplete representations has never been studied. The closest work we found considers a problem of multi-task regression~\cite{chen2010graph}. The final paper had been published under a different title~\cite{chen2010efficient} and proposes a new method based on the approximation of the fused-LASSO TV~penalty by a smooth convex function as described in~\cite{nesterov2005smooth}.


\section{Conclusion and Perspectives}
\label{sec:conclusion}
This paper has shown the efficiency of the proposed \XX~based on a split Bregman approach, 
in order to achieve the sparse structured approximation of multi-dimensional signals, under general conditions. 
Specifically, the extensive validation has considered different regimes in terms of the signal complexity and dynamicity (number of patterns simultaneously involved and average duration thereof), and it has established a {\em relative competence map} of the proposed \XX\ approach comparatively to the state of the art. 
Further work will apply the approach to the motivating application domain, namely the representation of EEG signals.


\pagebreak

\bibliographystyle{templates/IEEEbib}
\small{
\bibliography{biblio}
}

\end{document}